\documentclass{article}

\usepackage{arxiv}

\usepackage[utf8]{inputenc} 
\usepackage[T1]{fontenc}    
\usepackage{hyperref}       
\usepackage{url}            
\usepackage{booktabs}       
\usepackage{amsfonts}       
\usepackage{nicefrac}       
\usepackage{microtype}      
\usepackage{lipsum}
\usepackage{graphicx}
\graphicspath{ {./images/} }
\usepackage{natbib}
\usepackage{amsmath}
\usepackage{float}

\title{Bayesian Additive Regression Trees (BART) in Food Authenticity: A Classification Approach to Food Fraud Detection}

\author{
  Mengxiang Zhu \\
  School of Mathematics and Statistics \\
  University College Dublin \\
  Belfield, Dublin 4, Ireland, D04V1W8 \\
  \texttt{mengxiang.zhu@ucdconnect.ie}
  \and
  Riccardo Rastelli \\
  School of Mathematics and Statistics \\
  University College Dublin \\
  Belfield, Dublin 4, Ireland, D04V1W8 \\
  \texttt{riccardo.rastelli@ucd.ie}
}

\begin{document}
\maketitle
\begin{abstract}
Feature engineering plays a critical role in handling hyperspectral data and is essential for identifying key wavelengths in food fraud detection. This study employs Bayesian Additive Regression Trees (BART), a flexible machine learning approach, to discriminate and classify samples of olive oil based on their level of purity. Leveraging its built-in variable selection mechanism, we employ BART to effectively identify the most representative spectral features and to capture the complex interactions among variables. We use network representation to illustrate our findings, highlighting the competitiveness of our proposed methodology. Results demonstrate that when principal component analysis is used for dimensionality reduction, BART outperforms state-of-the-art models, achieving a classification accuracy of 96.8\% under default settings, which further improves to 97.2\% after hyperparameter tuning. If we leverage a variable selection procedure within BART, the model achieves perfect classification performance on this dataset, improving upon previous optimal results both in terms of accuracy and interpretability. Our results demonstrate that three key wavelengths, 1160.71 nm, 1328.57 nm, and 1389.29 nm, play a central role in discriminating the olive oil samples, thus highlighting an application of our methodology in the context of food quality. Further analysis reveals that these variables do not function independently but rather interact synergistically to achieve accurate classification, and improved detection speed.

\end{abstract}

\section{Introduction}
Extra virgin olive oil (EVOO) is one of the most frequently adulterated food products worldwide, representing one of the most persistent, economically motivated, and challenging forms of food fraud \citep{malavi2024,charlebois2016}. In recent years, Near Infrared HyperSpectral Imaging (NIR-HSI) has emerged as a promising technique for food quality evaluation and adulteration detection due to its non-destructive nature, high speed, and excellent accuracy. Unlike conventional near-infrared spectroscopy, HSI can simultaneously acquire spectral and spatial information \citep{cappelli2024,morabit2025}, allowing the identification and analysis of sample composition through its unique spectral characteristics, commonly referred to as spectral fingerprints \citep{siche2016}. However, the high-dimensional nature of HSI data presents significant analytical challenges. A typical dataset contains hundreds of spectral bands, many of which may be redundant, noisy, or weakly correlated with the target attributes. This complexity negatively affects model generalization, computational efficiency, and interpretability. As a result, the integration of statistical and machine learning methods has become essential for improving the accuracy, robustness, and scalability of predictive models when handling complex and heterogeneous spectral data \citep{gbashi2024,morabit2025}. In particular, feature selection plays a key role in identifying a minimal yet informative subset of wavelengths while maintaining model performance \citep{kamruzzaman2016,mendez2019}.

Although Malavi et al. (2024) \cite{malavi2024} employed various machine learning models to classify olive oil samples based on near-infrared hyperspectral imaging  data, their study primarily focused on comparing the predictive performance of the models, with limited attention given to variable interpretability and the identification of key wavelengths. In fact, feature selection and dimensionality reduction remain underexplored areas in the current literature \citep{ordoudi2023}. Identifying important wavelengths can provide valuable insights into the interactions among chemical structures in fats and oils \citep{mendez2019}. To address these gaps, this study builds upon the work of Malavi et al. (2024) \cite{malavi2024} by considering Bayesian Additive Regression Trees (BART) as an interpretable and probabilistically grounded framework for high-dimensional spectral data analysis. BART is a flexible ensemble learning algorithm that not only delivers highly accurate regression and classification but also supports built-in variable selection through posterior inclusion probabilities. Compared with conventional approaches, BART effectively captures nonlinear relationships and interactions among variables and provides quantitative uncertainty estimation for predictions, making it both highly practical and interpretable.

Based on this, our study aims to make the following contributions: (1) to employ BART as an advanced tool for EVOO adulteration detection, thereby expanding both the methodological toolkit for food authentication and the application scope of BART in high-dimensional spectral data analysis. (2) to compare BART with state of the art methods and demonstrate its superior performance in this dataset. (3) to select key predictors by identifying key wavelengths and to explore and illustrate their potential interactions using network analysis tools. This helps reduce the need for manual feature engineering by researchers and supports the design of low-cost, portable multispectral sensors for rapid and accurate on-site detection.

\section{Literature review}
\label{sec:headings}
Recent studies on olive oil adulteration detection have increasingly focused on achieving non-destructive, rapid, high-throughput, and highly sensitive analytical solutions. To address the limitations of traditional methods, namely high cost, time consumption, and complexity, researchers have explored the integration of spectroscopic techniques with machine learning algorithms to build high-performance adulteration classification and quantification models.

In the field of spectroscopy-based methods, various techniques such as visible-near infrared spectroscopy, Raman spectroscopy, fluorescence spectroscopy, and HSI have been employed to capture the physicochemical signatures of olive oils. For instance, Malavi et al. (2024) \cite{malavi2024} combined NIR-HSI with machine learning classifiers, demonstrating that artificial neural networks and partial least squares-discriminant analysis could achieve near-perfect classification performance (accuracy between 99.2–100\%) even for samples with just a subtle adulteration. Liang et al. (2025) \cite{liang2025} proposed a Raman spectroscopy framework integrated with a deep learning, achieving 99.3\% accuracy for identifying adulterant types while simultaneously estimating adulteration concentrations, thus fulfilling both qualitative and quantitative objectives. Similarly, Stavrakakis et al. (2022) \cite{stavrakakis2022} developed a multivariate statistical model based on visible absorption and fluorescence spectroscopy, capable of simultaneously predicting both adulterant type and concentration without any sample pretreatment, achieving over 90\% classification accuracy even for adulteration levels below 10\%. In another approach, Lu et al. (2023) \cite{lu2023} utilized pigment-based chromatographic fingerprints and a support vector machine approach, successfully detecting adulteration down to 1\% levels, suggesting pigment composition as a promising marker for fraud detection.

Beyond traditional spectral data, image-based modeling methods have emerged as powerful alternatives. Pradana-Lopez et al. (2022) \cite{pradana2022} trained convolutional neural networks on over 300,000 microscopy images of oil droplets, reaching a classification accuracy exceeding 96\%, highlighting the scalability and precision of image-driven approaches. Vega-Márquez et al. (2019) \cite{vega2019} applied similar methods to differentiate between EVOO, virgin olive oil (VOO), and lampante olive oil (LOO), achieving 82.8\% accuracy on independent test sets. Magdas et al. (2025) \cite{magdas2025} further advanced in this direction by extracting deep features from thermal and spectroscopic images using convolutional neural networks, and integrating them with other models for both adulterant type classification and concentration regression. Even under low adulteration levels (2.5–10\%), these models maintained high sensitivity and predictive performance, outperforming traditional statistical techniques. To address challenges such as unknown adulterants and imbalanced sample distributions, one-class SVMs and synthetic outlier generation strategies were also introduced to enhance model robustness and generalizability.

In summary, current research on olive oil adulteration detection is shifting from single-modality analysis toward multi-modal, multi-task, end-to-end intelligent systems. The deep integration of spectral and image-based data with AI models not only significantly enhances the detection of trace-level adulteration and complex mixtures, but also lays a solid foundation for building deployable, high-throughput, and scalable quality control solutions for the olive oil industry. This study focuses on hyperspectral data acquired through a cost-effective NIR-HSI technique and employs BART on this high dimensional dataset, incorporating two feature selection methods for dimensionality reduction and network visualization tools to explore variable interactions. The model’s strong predictive performance and efficiency highlight the practical potential of combining these approaches for oil food fraud detection.

\section{Data}
\label{sec:headings}

The dataset used in this study was collected from Malavi et al.(2024) \cite{malavi2024} and it is publicly available\footnote{\url{https://github.com/DNMalavi/NIR-HSI-ML-for-EVOO-Fraud-Detection}}.

\subsection{Data description}

The dataset consists of 1,995 near-infrared hyperspectral samples, covering 224 wavelengths ranging from 900 to 1700nm. It is designed for detecting adulteration in EVOO. The original samples include 36 EVOO from Spain, Italy, and Greece, along with 11 hazelnut oils, 11 olive pomace oils, and 6 refined olive oils samples. A subset of these oils was used to create adulterated mixtures at seven concentration levels (0\%, 1\%, 5\%, 10\%, 20\%, 40\%, and 100\%, of non-EVOO, by weight). Each mixture was prepared in triplicate to ensure measurement consistency, resulting in the complete dataset.

To enhance model performance, several spectral preprocessing techniques were applied, including normalization, standard normal variate transformation, Savitzky–Golay smoothing \citep{brown2000}, and derivative calculations. These steps reduce noise while enhancing spectral resolution, making the data more suitable for the statistical models, including BART. We aggregate the different mixtures into three classes: class 1 is made of just pure EVOO (0\% adulteration), class 2 is composed of partially adulterated EVOO (1\%, 5\%, 10\%, 20\%, 40\% adulteration), and class 3 only contains fully non-EVOO oils (100\% adulteration). We find that this distinction into three classes allows us to clearly separate pure oils (class 1 and 3) from adulterated samples (class 2), and so the characteristics of predictors may be ideal for discriminating the classes. A three-class problem can be more informative than a binary classification between pure and impure EVOO, however it also poses additional modeling challenges as some of the methods used are intended for binary classification problems. We note that other ways of aggregating the responses can also be considered. However, we can reasonably expect that this would not affect the quality of the results, since, as is shown in a later section, the accuracy obtained for some of the methods is in agreement with previous literature.

\subsection{Exploratory data analysis}
\begin{figure}
  \centering
  \includegraphics[width=0.5\textwidth]{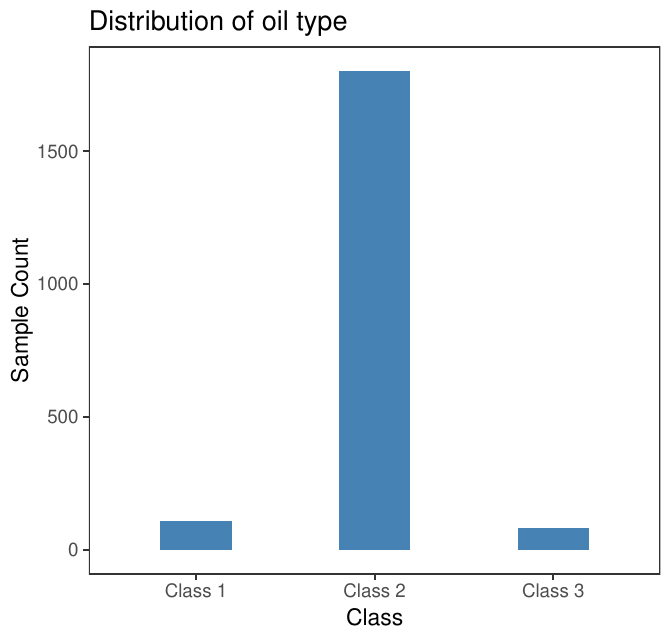}
  \caption{Response variable distribution plot. The x-axis represents the categories, where Class 1 denotes Pure EVOO, Class 2 denotes Partially adulterated EVOO including 1\%, 5\%, 10\%, 20\%, 40\% adulteration, Class 3 denotes non-EVOO with 100\% adulteration. The y-axis represents the count of samples in each category.}
  \label{fig:fig1}
\end{figure}

To facilitate the interpretability of the subsequent classification results, this section conducts a comprehensive exploratory data analysis to understand the structure, distribution, and potential challenges within the dataset. The primary aim of this task is to examine the distribution of the response variable, visually analyze the classes, and summarize the spectral predictors. This provides an initial assessment of signal quality, redundancy, and the discriminative potential across the wavelength ranges. Such insights are instrumental in guiding preprocessing decisions, developing potential dimensionality reduction strategies, and ultimately enhancing the model’s generalization capabilities.
Figure~\ref{fig:fig1} shows that the dataset exhibits a highly imbalanced distribution across the three classes. Class 2, which encompasses the majority of adulterated samples at various concentrations, contains as many as 1,803 samples, whereas Class 1 and Class 3 include only 108 and 84 samples, respectively. During model fitting, the overwhelming size of Class 2 may affect the model's learning process, leading it to focus disproportionately on patterns from this class and thereby weakening its ability to accurately classify Class 1 and Class 3. Therefore, employing additional techniques such as principal component analysis (PCA) or variable selection to address this imbalance and extract more relevant wavelength information is crucial for enhancing the classification and predictive performance for Class 1 and Class 3. This constitutes a necessary step toward optimizing the model and ensuring fairness.

\begin{figure}
  \centering
  \includegraphics[width=0.8\textwidth]{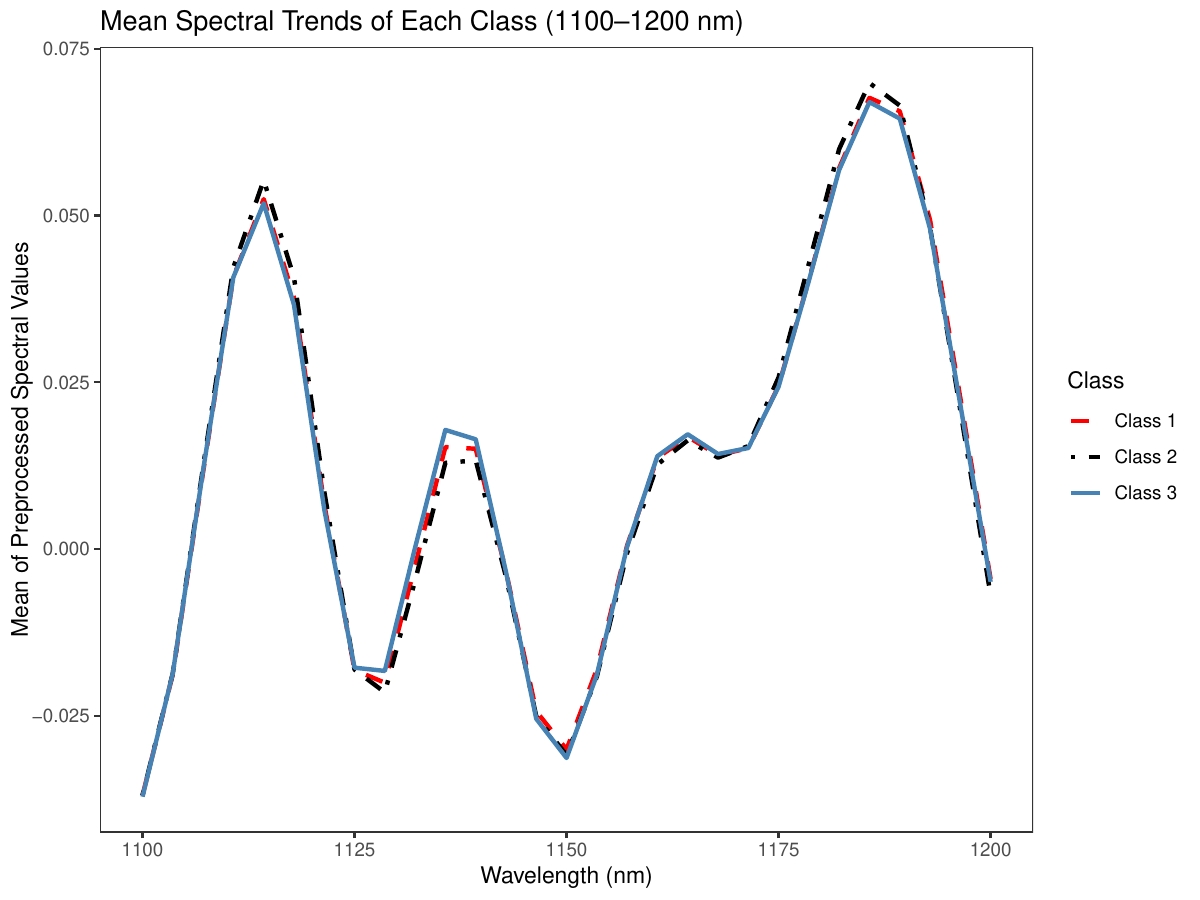}
  \caption{Average spectral profiles of the three response classes (Class 1 denotes pure EVOO, Class 2 denotes partially adulterated EVOO and Class 3 denotes fully adulterated oils) within the wavelength range of 1100-1200 nm. The x-axis represents the wavelengths in nanometers, and the y-axis shows the mean of the preprocessed spectral values for each class at each wavelength. These averaged curves illustrate partial spectral patterns of different classes and help reveal wavelength regions with potential discriminative power for classification.}
  \label{fig:fig2}
\end{figure}

Due to the high chemical similarity between adulterated oils and pure EVOO, their near-infrared spectral curves exhibit a strong degree of consistency. However, noticeable differences still emerge in certain wavelength regions. As shown in Figure~\ref{fig:fig2}, within the 1100–1200 nm range, the average spectral curves of the three oil classes largely overlap, but visible separations appear around several peak positions, indicating that specific wavelength intervals within this region may serve as discriminative features for class separation.

The high degree of overlap in the majority of the wavelength range suggests that classification may be challenging unless the model is trained to focus on these specific discriminative regions. Therefore, identifying and leveraging these key wavelength intervals could improve the model's performance by enhancing its ability to differentiate between the pure EVOO, partially adulterated EVOO, and fully adulterated oils. Further exploration of these spectral regions may reveal more nuanced patterns, which could be critical for refining feature selection strategies and enhancing classification accuracy.
\newpage

\section{BART models}
\label{sec:headings}

Based on the insights gained from the exploratory data analysis, this study introduces BART, a novel machine learning classifier, for discriminating pure EVOO, partially adulterated EVOO, and pure adulterants. Our approach aims not only at enhancing the accuracy and robustness of classification models in distinguishing between pure and adulterated olive oils, but also it enables the identification of critical wavelength features through the feature selection framework that is embedded in BART. The framework incorporates regularization to prevent overfitting and to effectively highlight the most informative spectral regions. Additionally, we propose some interpretable illustrations of the interactions between predictors using network analysis tools.

\subsection{Bayesian Additive Regression Trees}

Bayesian Additive Regression Trees were originally introduced by Chipman et al.(2010)\cite{chipman2010}. This is a Bayesian non-parametric statistical algorithm that employs an ensemble of trees. It combines multiple shallow decision trees as weak learners, with each tree contributing a small portion to the overall approximation of an unknown target function. The final prediction is obtained by summing the outputs of all individual trees, enabling BART to effectively capture complex nonlinear relationships and interaction effects among predictors. It is suitable for both regression and classification tasks. Given a response vector $Y$ of length $n$ and a covariate matrix $X$ with $d$ predictors, the BART model can be expressed as:

\begin{equation}
Y = \sum_{j=1}^{m} g(X; T_j, M_j) + \varepsilon, \quad \varepsilon \sim N(0, \sigma^2).
\end{equation}

where $ g(X; T_j, M_j) $ is a function which calculates the individual contribution of each tree $j$ of $J$ total trees. $M_j$ specifies the terminal node parameters associated with the $j$th tree $T_j$. The residuals, $\varepsilon$, are assumed to be normally distributed with mean 0 and variance $\sigma^2$.

\begin{figure}[H]
  \centering
  \includegraphics[width=0.3\textwidth]{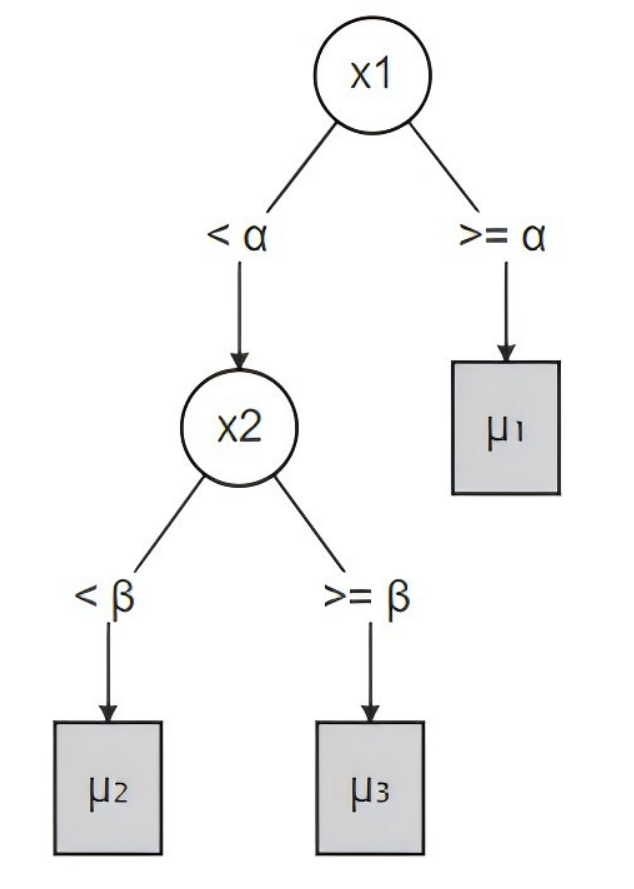}
  \caption{A sample decision tree plot, where $ x_1 $ and $ x_2 $ represent the covariates, $ \alpha $ and $ \beta $ represent split value and $ \mu_1 $, $ \mu_2 $, $ \mu_3 $ represent the leaf node value.}
  \label{fig:fig3}
\end{figure}

The function $ g(x; T_j, M_j) $ maps an input $ x $ to a terminal node value $ \mu_{j,k} \in M_j $, depending on the partition defined by the tree $ T_j $. The function $ g(x; T_j, M_j) $ may depend on a single variable, or it may involve an interaction effect by involving multiple variables. A graphical representation of $ g(x; T_j, M_j) $ is given in Figure~\ref{fig:fig3}.

As a fully Bayesian model, BART requires suitable prior distributions for the tree structures $ T_j $, terminal node parameters $ M_j $, and the error variance $ \sigma^2 $. Posterior inference is performed using a Markov Chain Monte Carlo (MCMC) backfitting algorithm, which iteratively updates the structure and parameters of each tree. Given the observed data $ y $, the model induces a posterior distribution:
\begin{equation}
p\bigl((T_1, M_1), \ldots, (T_m, M_m), \sigma \mid y\bigr).
\end{equation}
which is targeted by the sampler. This allows BART to provide not only point estimates but, potentially, also credible intervals, offering a built-in mechanism for quantifying uncertainty in predictions.

\subsection{Multinomial BART for categorical outcomes}

Although BART has demonstrated strong performance in both regression and classification tasks, its early implementations were primarily limited to binary outcomes, which precluded its applicability to multiclass classification problems frequently encountered in real-world datasets. To address this limitation, Sparapani et al. (2021) \cite{sparapani2021} extended the original BART model proposed by Chipman et al. (2010) \cite{chipman2010} and developed the \texttt{R} package \texttt{BART}, which incorporates a function named \texttt{mbart} to support multinomial classification. This implementation provides options for both \textit{probit} and \textit{logistic} link functions.

In this study, the multinomial BART approach based on the probit link will be employed. Compared to the logistic link, the probit formulation introduces latent Gaussian variables, which typically enable more efficient posterior sampling, improved computational performance, and better convergence stability, while maintaining competitive predictive accuracy and interpretability. As a result, the probit-based BART model is particularly well suited to multiclass classification tasks in high-dimensional feature spaces, such as in hyperspectral food fraud detection, where computational cost and convergence are critical considerations.

Given that the classification task in this study involves three classes, the multinomial BART framework assumes $K=3$ mutually exclusive classes, each represented by a binary indicator variable. To predict the categories, the model first computes the conditional probabilities for each category:

\begin{align}
p_{i1} &= P(y_{i1}=1) \\
p_{i2} &= P(y_{i2}=1 \mid y_{i1}=0) \\
p_{i3} &= P(y_{i3}=1 \mid y_{i1}=0, y_{i2}=0).
\end{align}

For each observation, the model estimates the conditional probability of membership in each class:

\begin{align}
\pi_{i1} &= p_{i1} \\
\pi_{i2} &= p_{i2} (1 - p_{i1}) \\
\pi_{i3} &= p_{i3} (1 - p_{i1}) (1 - p_{i2}).
\end{align}

This formulation ensures that the predicted class probabilities for each observation sum to one, that is $\pi_{i1} + \pi_{i2} + \pi_{i3} = 1$, thereby supporting valid probabilistic classification. The final predicted class is assigned based on the class with the highest posterior probability.

This can be understood as the model fitting a sequence of binary sub-models, each predicting whether a given observation belongs to a specific class (indicator = 1) or not (indicator = 0). These conditional probabilities are then normalized to obtain unconditional class probabilities for each observation.

The hyperspectral dataset consists of olive oil samples characterized by 224 spectral features, and the class boundaries are not linearly separable. Therefore, the application of multinomial BART not only improves classification accuracy but also offers the potential to capture subtle sample-level variation and structural patterns.

\section{Other methods}
\label{sec:headings}

To ensure a consistent baseline across different models, this study employs Principal Component Analysis (PCA) as an initial preprocessing step for dimensionality reduction prior to classification. Principal Components (PCs) that capture the majority of the data variance are selected and used as predictors across four models: Multinomial BART, Random Forest (RF), Extreme Gradient Boosting (XGBoost), and Multinomial Logistic Regression (MLR). By using PCA not only we enhance model comparability, but also the risks of multicollinearity and overfitting are effectively mitigated. Furthermore, PCA provides a standardized foundation for subsequent feature selection analyses, facilitating a direct comparison between PCA-based and model-driven variable selection strategies.

All models are evaluated using a unified cross-validation framework to ensure fair comparison. The final model performance is assessed and compared across all four methods, enabling a comprehensive evaluation of the predictive effectiveness and robustness of Multinomial BART relative to conventional machine learning and statistical classification approaches.

\subsection{Principle Component Analysis}

\begin{figure}[H]
  \centering
  \includegraphics[width=0.6\textwidth]{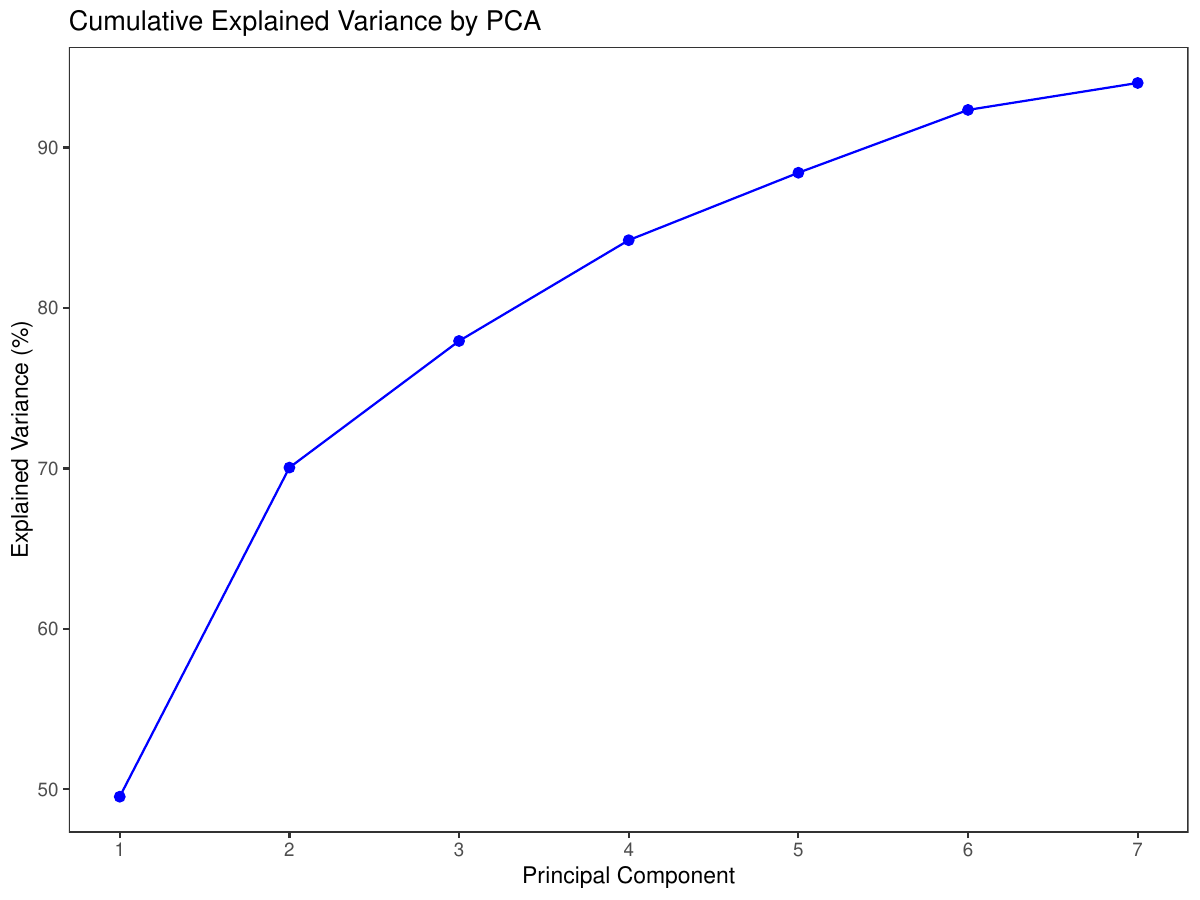}
  \caption{PCs cumulative explained variance plot to visualize how much total variance each principal component, and their cumulative sum, explains in the dataset. X-axis represents the index of principal components and Y-axis represents the cumulative variance explained percentage.}
  \label{fig:fig4}
\end{figure}

In this section, we describe how we employed PCA to reduce the 224-dimensional feature space. PCA transforms the original high-dimensional features into a set of orthogonal PCs, which are uncorrelated with each other and capture the directions of maximum variance in the data. The cumulative proportion of total variance explained by the PCs is illustrated in Figure \ref{fig:fig4}.

In this study, PCA was conducted using all spectral wavelength variables, and up to 7 components were considered in the model. Each PC is a linear combination of the original variables and captures a distinct direction of variance in the feature space. Importantly, the components are orthogonal to each other, meaning they are mutually uncorrelated. This property allows PCA to effectively address both multicollinearity and dimensionality reduction in high-dimensional data.

As shown in Figure~\ref{fig:fig4}, the first two principal components (PC1 with 49.54\% and PC2 with 20.51\%) together account for approximately 70\% of the total variance in the original feature set. In contrast, the explanatory power of subsequent components drops off sharply. For example, the third PC accounts for only 6.28\% of the variance. Therefore, to strike a balance between preserving meaningful information and minimizing dimensional complexity, only PC1 and PC2 were selected as predictors in the classification models. These two components were then used as inputs to the Multinomial BART model, as well as to three baseline models to evaluate and compare their classification performance.

\subsection{Multinomial BART}

The multinomial BART model was first fit using the training data and default hyperparameters. The default configuration includes: 200 for the number of trees, $k=2$, $power=2$, $base=0.95$ for the trees' prior parameters. To improve model performance, a grid search with 5-fold cross-validation was performed on the training set. The goal of this cross-validation procedure was to identify the optimal combination of hyperparameters that minimize classification error. Specifically, four key hyperparameters were tuned: number of trees, tree prior $k$, tree prior power and tree prior base.

A $log$ loss was used as the objective metric for tuning, as it directly quantifies the accuracy and confidence of probabilistic predictions in multiclass settings. The average $log$ loss across all folds was computed for each parameter setting, and the configuration with the lowest mean $log$ loss was selected as the final model setup.

\subsection{Random Forest}

Random Forest (RF) is an ensemble learning algorithm that builds multiple decision trees and aggregates their results to improve classification accuracy and reduce overfitting. It utilizes two main techniques: bootstrap aggregation (bagging) and random feature selection, which enhance model robustness and performance \citep{breiman2001}.

Since only the first two principal components (PC1 and PC2) were used as input features, the strategy of tuning \textit{mtry} (the number of variables considered at each split) was not applicable. Instead, a grid search was conducted to optimize two key hyperparameters: the number of trees (\textit{ntree}) and the maximum number of terminal nodes per tree (\textit{maxnodes}). Model performance was internally validated using out-of-bag (OOB) error estimation and further refined via 5-fold cross-validation. Final evaluation was performed on an independent test set to assess predictive generalizability.

\subsection{Extreme Gradient Boosting}

Extreme Gradient Boosting (XGBoost) is a scalable, tree-based gradient boosting algorithm that has become one of the most effective and widely used machine learning methods for classification tasks, particularly in structured data problems. It sequentially builds decision trees by optimizing a differentiable loss function through gradient descent, and includes several regularization techniques to reduce overfitting. The model was trained using the \textit{multi:softprob} objective, which outputs class probabilities for multiclass classification. The evaluation metric was set to multiclass log-loss.

To determine the optimal number of boosting rounds, 5-fold cross-validation was performed with early stopping set to 10 rounds. The model with the lowest cross-validated log-loss was selected, and then retrained on the training set using the optimal number of rounds. Predictions were made on the external test set, and the final class label for each sample was assigned based on the class with the highest predicted probability. Model accuracy was then calculated as the proportion of correctly predicted labels in the test set.

\subsection{Multinomial Logistic Regression}

Multinomial Logistic Regression (MLR) is an extension of binary logistic regression that enables the modeling of response variables with more than two discrete, non-ordered categories. MLR computes the probabilities of each class relative to a designated reference (baseline) category. For each non-reference class, the model estimates a distinct set of coefficients, thereby capturing how the predictor variables influence the odds of an observation being classified into each specific category. Based on the input features, the model outputs a probability distribution across all classes, and the final classification is assigned to the category with the highest predicted probability.

In this study, The MRL model was trained using 5-fold cross-validation with stratified sampling to maintain balanced class distributions across folds. Hyperparameter tuning was conducted via grid search over a range of regularization strength (\textit{decay}) values. To ensure model convergence, training was allowed to run for up to 1000 iterations. Upon training completion, the model’s performance was assessed using predictions on the test set, with a confusion matrix constructed to evaluate classification accuracy and class-wise performance.

\section{Variable selection in BART}

Variable selection in BART aims to identify and retain the most informative predictors while eliminating irrelevant or redundant ones. This process helps reduce model complexity, enhance interpretability, and potentially improve predictive accuracy, particularly in high-dimensional feature spaces. In the context of spectral data, it further facilitates the identification of key wavelength regions that contribute most to classification performance. Two main approaches were employed to implement variable selection within the BART framework, as shown in the subsections below.

\subsection{Variable usage frequency}

The first method for evaluating variable importance in BART is to calculate the percentage with which each variable is selected for tree splits. BART performs posterior sampling via MCMC, and each iteration generates a set of regression trees. At each split node, a variable is selected from all available predictors as the splitting criterion. The frequency with which a variable is chosen at split nodes reflects its importance. This method uses the \texttt{varcount} function to extract, for each MCMC iteration, the number of times each variable appears across all trees. Each row corresponds to one MCMC sample, and each column corresponds to one variable. Since the total number of splits may vary across samples, each row is first normalized to obtain the relative frequency of variable usage in that iteration:

\begin{equation}
f_{ij} = \frac{v_{ij}}{\sum_{k=1}^{p} v_{ik}}
\end{equation}

Where $ v_{ij} $ is the number of times variable $ j $ is used as a split in the $ i $-th MCMC sample, $ v_{ik} $ is the number of times the $ k $-th variable is selected in the $ i $-th sample, and $ p $ is the total number of variables, so that $ \sum_{k=1}^{p} v_{ik} $ is the total number of splits across all variables for the $ i $-th iteration. The average usage frequency overall $ N $ iteration is then calculated as:

\begin{equation}
\overline{f}_j = \frac{1}{N} \sum_{i=1}^{N} f_{ij}.
\end{equation}

\subsection{Sparse prior-based frequency}

The second method enables variable selection by setting \texttt{sparse = TRUE}, which introduces a sparse prior. This prior assumes that only a small subset of variables are truly important and assigns a Dirichlet prior to their selection probabilities $ s_j $, with the sparsity level controlled by a Beta prior. During MCMC sampling, $ s_j $ is updated iteratively: important variables see their selection probabilities increase, while unimportant ones decrease. The sparse Dirichlet framework ensures that the number of relevant variables is regularized and thus that the model remains parsimonious. This directly affects subsequent tree-splitting decisions, making important variables more likely to be chosen and noise variables rarely selected. The average posterior selection probability for variable $ j $ is calculated as:

\begin{equation}
\overline{S}_j = \frac{1}{N} \sum_{i=1}^{N} s_{ij}.
\end{equation}
where $ s_{ij} $ is the posterior selection probability of variable $ j $ in iteration $ i $, and $ N $ is the total number of MCMC iterations.

In practice, these average probabilities can be directly extracted from the model output using \texttt{\$varprob.mean}. By comparing $ \overline{S}_j $ with the threshold $ 1/p $, where $ p $ is the total number of variables, variables exceeding the threshold are identified as important for prediction.

\section{Statistical analysis}
\subsection{Training and testing of models}

In this study, four machine learning models were constructed and fitted using \texttt{RStudio Posit} (version 4.3.2), namely: BART, Random Forest, XGBoost, and MLR. The primary \texttt{R} packages used included \texttt{BART}, \texttt{randomForest}, \texttt{xgboost}, and \texttt{caret}. During model training, 5-fold cross-validation was implemented using the \texttt{trainControl} function from the \texttt{caret} package, and model fitting was carried out with the \texttt{train} function.

To address class imbalance in the dataset, Synthetic Minority Over-sampling Technique (SMOTE) was applied in the training folds for the Random Forest, XGBoost, and Multinomial Logistic Regression models. It is important to note that SMOTE was applied only to the training folds during cross-validation; the validation folds and the final test set retained their original distributions to ensure objective and unbiased performance evaluation.

Before training, the dataset was split into a calibration set and a test set. The calibration set consisted of 1,390 samples (approximately 70\% of the total), used for model training and hyperparameter tuning. The remaining 599 samples (approximately 30\%) formed the test set, used to assess the generalization performance of the final models on unseen data.

To obtain optimal hyperparameter configurations, either grid search or random search strategies were employed to explore predefined parameter spaces. These tuning procedures were integrated with 5-fold cross-validation, whereby each hyperparameter setting was evaluated iteratively across the calibration set. The parameter configuration that achieved the best average performance on the validation folds was selected for the final model. This approach effectively reduces the dependency on a specific data split and enhances the model’s stability and robustness.

\subsection{Evaluation of model performance}

The performance of each binary machine learning classification algorithm was evaluated using metrics such as accuracy, sensitivity, precision, specificity, F1 score, and Matthews Correlation Coefficient (MCC), calculated from confusion matrices, as shown in the following equations \citep{desantana2018,vanroy2018}.

\begin{equation}
Accuracy = \frac{TP + TN}{TP + TN + FP + FN},
\end{equation}

\begin{equation}
Sensitivity = \frac{TP}{TP + FN},
\end{equation}

\begin{equation}
Specificity = \frac{TN}{TN + FP},
\end{equation}

\begin{equation}
Precision = \frac{TP}{TP + FP},
\end{equation}

\begin{equation}
F1 \ Score = 2 \times \frac{Precision \times Sensitivity}{Precision + Sensitivity},
\end{equation}

\begin{equation}
MCC = \frac{(TP \times TN - FP \times FN)}{\sqrt{(TP + FN)(TP + FP)(TN + FN)(TN + FP)}}.
\end{equation}

Where TP = true positives, TN = true negatives, FN = false negatives, and FP = false positives. Accuracy measures the overall proportion of correct predictions, while sensitivity and specificity assess the model’s ability to correctly identify positive and negative samples, respectively. Sensitivity (recall or true positive rate) reflects the proportion of actual positive cases that are correctly predicted by the model, indicating how effectively the model captures the positive class. Specificity (true negative rate), measures the proportion of actual negative cases correctly identified, representing the model’s capacity to avoid false positives. Precision focuses on the accuracy of positive predictions, and the F1 score balances precision and sensitivity, which is particularly useful for imbalanced datasets. The Matthews Correlation Coefficient (MCC), a robust metric for assessing agreement between predicted and actual classes, was also used due to the numerical imbalance of sample groups. MCC values range from -1 to 1, where -1 indicates complete disagreement, 0 indicates random chance, and 1 indicates perfect agreement between predicted and actual classes \citep{chicco2023}.

\section{Results}

\subsection{Comparison of classification algorithms}

This modeling phase was designed to investigate two key research questions: (1) model performance comparison. We evaluate the performance of BART against other tree ensemble methods (RF, XGBoost), as well as MLR, a standard statistical method for multi-class classification, to understand their relative strengths in this context. (2) classification difficulties in identifying Class 1 and Class 3. Given the severe imbalance in the response variable, characterized by a predominance of Class 2 samples, the classification models exhibit an inherent bias towards this majority class, thereby exacerbating the challenge of accurately discriminating between Class 1 and Class 3. Consequently, the analysis primarily focuses on evaluating the models’ classification performance on these two minority classes.

\subsubsection{Performance metrics}

Based on cross-validation, the optimal hyperparameters for each model were identified as follows: BART: $ntree=50$, $k=1$, $power=1.5$, $base=0.95$, with average cross-validation loss of 0.0657; RF: $ntrees=100$, $maxnodes=20$; XGBoost: $nrounds=20$ (optimal number of boosting iterations); MLR: $decay=1e-4$ (L2 regularization strength). Using the tuned parameters, all models were trained on the training dataset and evaluated on a hold-out test set. Table~\ref{tab:table1} summarizes key performance metrics.
\newpage

Table 1: shows the classification performance of BART with default hyperparameters (BART), and its cross-validated variant (BART\_CV), RF cross-validation variant (RF\_CV), XGBoost cross-validation variant (XGBoost\_CV), and MLR cross-validated variant (MLR\_CV), evaluated across five key metrics: accuracy, MCC, sensitivity, specificity, and F1 score. Among them, BART and BART\_CV serve as the baseline models for the comparisons.
\vspace{5em}

\begin{table}
 \caption{Performance metrics}
  \centering
  \begin{tabular}{l c c c c c}
    \toprule
    \cmidrule(r){1-2}
    Model & Accuracy & MCC & Sensitivity & Specificity & F1 Score \\
    \midrule
    BART & 0.968 & 0.784 & 0.798 & 0.989 & 0.785 \\
    BART\_CV & \textbf{0.972} & 0.804 & 0.820 & \textbf{0.990} & \textbf{0.812} \\
    RF\_CV & 0.971 & \textbf{0.848} & \textbf{0.821} & \textbf{0.990} & 0.806 \\
    XGBoost\_CV & 0.968 & 0.828 & 0.797 & 0.989 & 0.791 \\
    MLR\_CV & 0.956 & 0.760 & 0.750 & 0.954 & 0.757 \\
    \bottomrule
  \end{tabular}
  \label{tab:table1}
  \end{table}

In terms of accuracy, BART demonstrated strong baseline performance even with default parameters, achieving 0.968, only marginally lower than RF (0.971), a mere 0.3\% difference. After cross-validation tuning, BART\_CV achieved the highest accuracy (0.972) among all models, indicating its excellent adaptability and learning capacity under optimized settings. The sensitivity results suggest that BART\_CV is highly capable of identifying positive instances, such as accurately distinguishing pure or fake olive oil, performing on par with RF\_CV and outperforming both XGBoost\_CV (0.797) and MLR\_CV (0.750). Regarding the specificity, all three tree-based models (BART\_CV, RF\_CV, XGBoost\_CV) performed exceptionally well, with BART\_CV and RF\_CV both achieving 0.990, reflecting strong ability to correctly identify negative cases and maintain low false-positive rates. While BART\_CV lags slightly behind RF\_CV in MCC, the gap is relatively notable compared to the differences in other metrics. Importantly, BART\_CV achieved the highest F1 Score (0.812) among all models, showcasing its effectiveness in balancing precision and recall across all three categories, which is particularly valuable in scenarios involving class imbalance or class overlap.

However, when viewed through the lenses of the MCC, BART\_CV appears to underperform slightly relative to RF\_CV. MCC, unlike accuracy or F1 score, is a more balanced metric that takes into account true and false positives and negatives across all classes, making it especially informative for multi-class classification problems with imbalanced datasets. The lower MCC score for BART\_CV (0.804) compared to RF\_CV (0.848) suggests that, despite its strong average performance, BART\_CV may exhibit more inconsistency or misclassification in specific categories, especially in borderline or ambiguous cases.

\subsubsection{Confusion matrix}
\begin{figure}[H]
  \centering
  \includegraphics[width=0.8\textwidth]{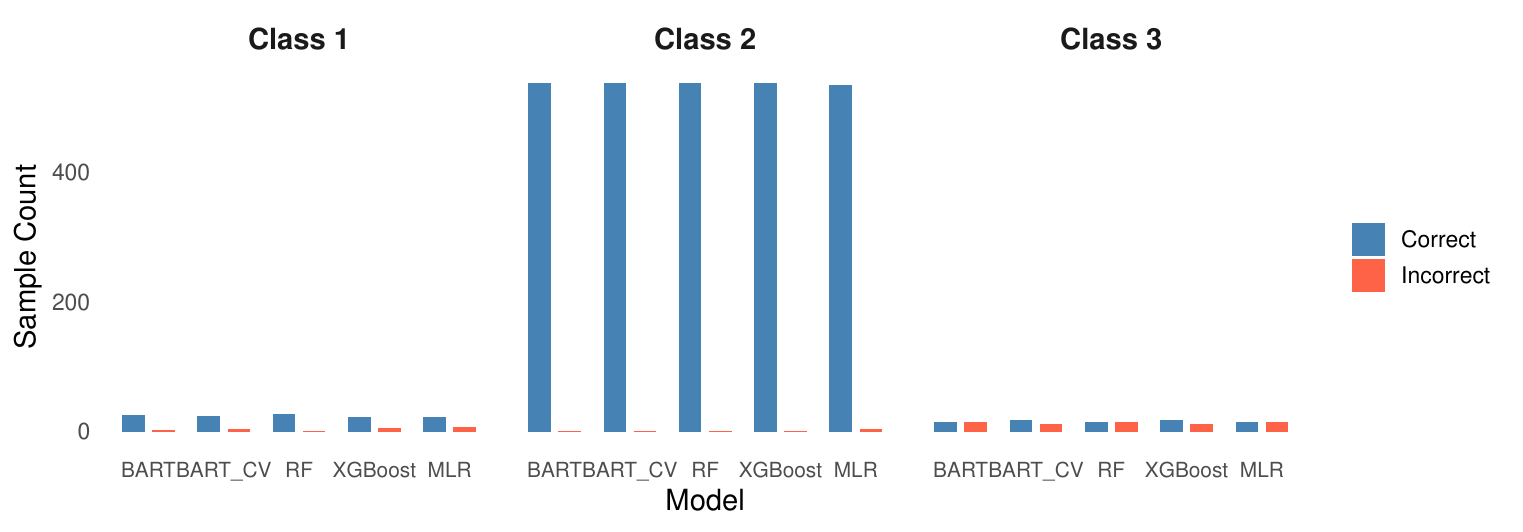}
  \caption{Bar plot comparing the number of correctly and incorrectly classified samples based on confusion matrix for each model across three classes. Each panel corresponds to a class, and bars are grouped by model type. Correct and incorrect classifications are distinguished by color.}
  \label{fig:fig5}
\end{figure}

A more detailed understanding of each model’s classification behavior can be gained by Figure~\ref{fig:fig5}, the bar plot created based on the confusion matrices. It highlights where and how each model succeeds or fails in specific class predictions.

The results show that all tree-based ensemble models showed excellent performance on Class 2, with each model misclassifying only one sample in this category. This indicates a strong capacity to capture the intermediate characteristics of partially adulterated olive oil. For Class 1, RF\_CV demonstrated the best identification ability, misclassifying only one sample, suggesting that RF\_CV has a distinct advantage in recognizing the authentic class compared to other models. However, all models struggled with Class 3. In particular, both BART and RF\_CV only correctly classified about 50\% of the samples in this class. This reflects the difficulty in distinguishing fully adulterated samples, which may exhibit overlapping spectral or statistical properties with Class 2 or even Class 1 in certain dimensions. In contrast, MLR showed generally lower classification performance across all three classes, failing to achieve standout results in any category. Its confusion matrix reveals a higher rate of misclassification in both Class 1 and Class 3, confirming its limitations in capturing non-linear or complex feature interactions inherent in this dataset. While BART and BART\_CV performed well overall, their difficulties with Class 3 suggest potential areas for improvement.

In conclusion, BART\_CV proves to be a highly competitive model, offering performance comparable to or exceeding that of state-of-the-art ensemble models like RF and XGBoost in most metrics. Although BART did not achieve the best performance in its default parameter setting, the results show that its performance was not substantially inferior to that of the cross-validated version. This suggests that even without hyperparameter tuning, BART can still deliver reasonably reliable results, making it a robust choice when tuning resources are limited or unavailable. However, its performance under default setting, particularly in handling imbalanced class distributions, still leaves room for improvement. To address this, the following sections will focus on feature selection from the full hyperspectral dataset under the default BART model. The aim is to explore whether variable selection can further enhance BART’s classification accuracy and robustness, especially in the context of imbalanced multi-class problems. This additional analysis will also help evaluate BART’s potential not only as a strong predictive model but also as a tool for identifying key discriminative features in high-dimensional settings.

\subsection{Variable selection using BART}

In this section, we further refine the results by considering two methods for BART variable selection. The aim is to evaluate the effectiveness of BART's internal variable selection mechanism and to identify the most representative spectral features. After the selection process, a reduced subset of variables is used to retrain and test the BART model, enabling the assessment of the accuracy and reliability of BART-based variable selection, as well as its impact on overall model performance. This process not only validates the effectiveness of BART's variable selection strategy, but also provides valuable insights into the key spectral characteristics that distinguish pure, partially adulterated, and fully adulterated olive oil samples.

\subsubsection{Variable usage frequency}

The first approach leverages the \texttt{varcount} output from the BART model, which records how often each variable is used in tree split decisions during each post–burn-in MCMC iteration. Each row of the \texttt{varcount} matrix corresponds to one MCMC iteration, while each column represents a specific input variable. To quantify the importance of each variable, the raw counts were normalized within each iteration to obtain per-iteration usage proportions, followed by averaging across all iterations to compute the posterior mean usage proportion for each variable. This metric reflects the average frequency with which a variable contributes to decision rules in the BART model, serving as a proxy for its predictive relevance.

\begin{figure}[H]
  \centering
  \includegraphics[width=0.6\textwidth]{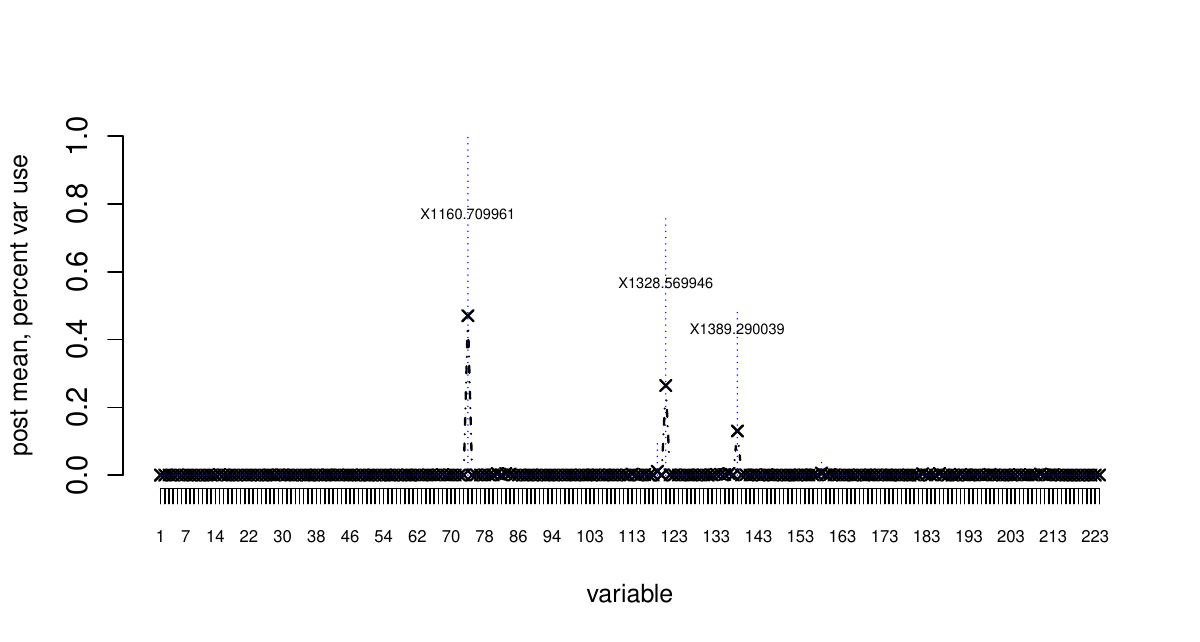}
  \caption{Percent of all variables used in the trees. The x-axis represents the variable indices, and the y-axis indicates the posterior mean proportion of times each variable was used in the tree-splitting process across post-burn-in MCMC iterations. Each point corresponds to a single variable's average usage frequency, while the blue dashed lines denote the 95\% credible intervals, reflecting the uncertainty in variable inclusion across posterior samples.}
  \label{fig:fig6}
\end{figure}

Based on these results, the top 5 most frequently used variables were identified. Importantly, from Figure~\ref{fig:fig6} and Table~\ref{tab:table2}, the top 3 variables are accounted for approximately 86\% of total variable usage, while the remaining 221 variables contributed only marginally and even close to zero. This highly concentrated distribution suggests that the BART model heavily relies on these few variables for decision-making, and that the vast majority of features provide little to no additional predictive value.

\begin{table}[H]
 \caption{Top 5 Important variables}
  \centering
  \begin{tabular}{l c c c c c}
    \toprule
    \cmidrule(r){1-2}
    Variable & X1160.709961 & X1328.569946 & X1389.290039 & X1321.430054 & X1460.709961 \\
    \midrule
    Mean Proportion & 0.4704 & 0.2645 & 0.1302 & 0.0108 & 0.0061 \\
    \bottomrule
  \end{tabular}
  \label{tab:table2}
  \end{table}

However, it is important to note that the credible intervals associated with these variable usage probabilities, particularly those derived from the posterior distribution across MCMC iterations, are relatively wide. In the case of X1160.71, the variable with the highest average usage, its credible interval spans a wide range. This uncertainty suggests that the variable selection results, while informative, should be interpreted with caution, especially when using BART for feature selection in high-dimensional settings. The wide intervals reflect instability in variable inclusion across posterior draws, possibly due to collinearity, redundant information, or data sparsity in certain regions of the feature space. Despite this uncertainty, the clear dominance of the top three variables in terms of average usage makes a strong case for their inclusion in a reduced-feature model.

\begin{figure}[H]
  \centering
  \includegraphics[width=0.6\textwidth]{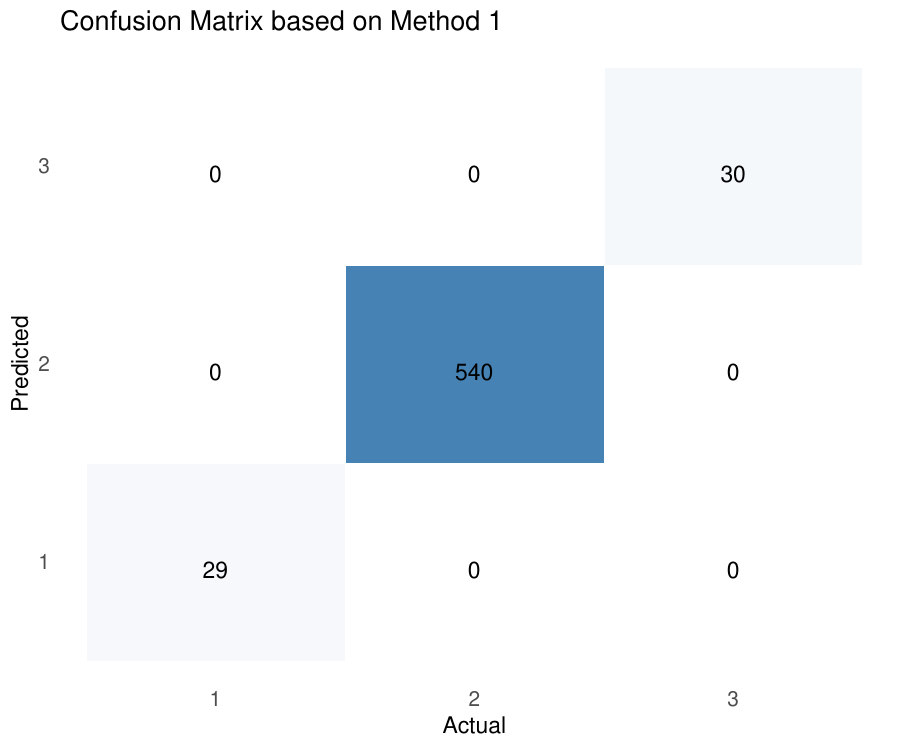}
  \caption{Confusion matrix heatmap for BART model in testing data with variables selected by method 1 (\texttt{varcount}).}
  \label{fig:fig7}
\end{figure}

Using the top three variables: X1160.709961, X1328.569946, and X1389.290039, which together account for approximately 86\% of total variable usage, the BART model was retrained and tested using only these predictors. The classification results on the test dataset are shown in Figure \ref{fig:fig7}.

As illustrated in Figure~\ref{fig:fig7}, the model achieved perfect classification across all three classes: no misclassifications occurred, and every test sample was correctly labeled. Consequently, all performance metrics, including accuracy, sensitivity, specificity, F1-score, and MCC, reached their maximum value of 1.0. These results demonstrate that the variable selection method based on BART’s internal usage frequency not only effectively reduces dimensionality, but also enhances model performance by isolating the most informative predictors. The outcome suggests that this approach allows the model to better capture discriminative patterns and structure in the data, resulting in more accurate and robust classification, even in a multi-class setting.

\subsubsection{Sparse prior-based frequency}

The second method applies class-specific variable selection using the BART model’s output under a sparse Dirichlet prior. In this framework, the model is trained as a series of binary classifiers, each focused on separating a particular class from the rest. The variable selection probabilities for each class are derived from the \texttt{\$varprob.mean} output, which reports the mean posterior probability that a given variable is used in the tree-splitting process across MCMC iterations.

As a result, the sample output matrix contains two rows per variable shown in Table~\ref{tab:table3}:

\begin{table}[H]
 \caption{Example of the mean probability that variables are selected during tree splitting}
  \centering
  \begin{tabular}{l c c c c c c}
    \toprule
    X900 & X903.570007 & X907.140015 & X910.710022 & X914.289978 & X917.859985 & X921.429993 \\
    \midrule
    0.0002735828 & 0.0001317964 & 0.0004553638 & 1.681566e-04 & 0.0010907357 & 0.0002562565 & 0.0002248531 \\
    0.0001741713 & 0.0001364988 & 0.0001934625 & 9.246172e-05 & 0.0003609426 & 0.0001201089 & 0.0007835871 \\
    \bottomrule
  \end{tabular}
  \label{tab:table3}
  \end{table}

Table 3: Samples of mean posterior probability that a given variable is used in the tree-splitting process across MCMC iterations. The first row represents the average selection probability for predicting Class 1 (pure EVOO). The second row reflects the variable importance in predicting Class 2 (partially adulterated), conditioned on the exclusion of Class 1.

\begin{figure}
  \centering
  \includegraphics[width=0.6\textwidth]{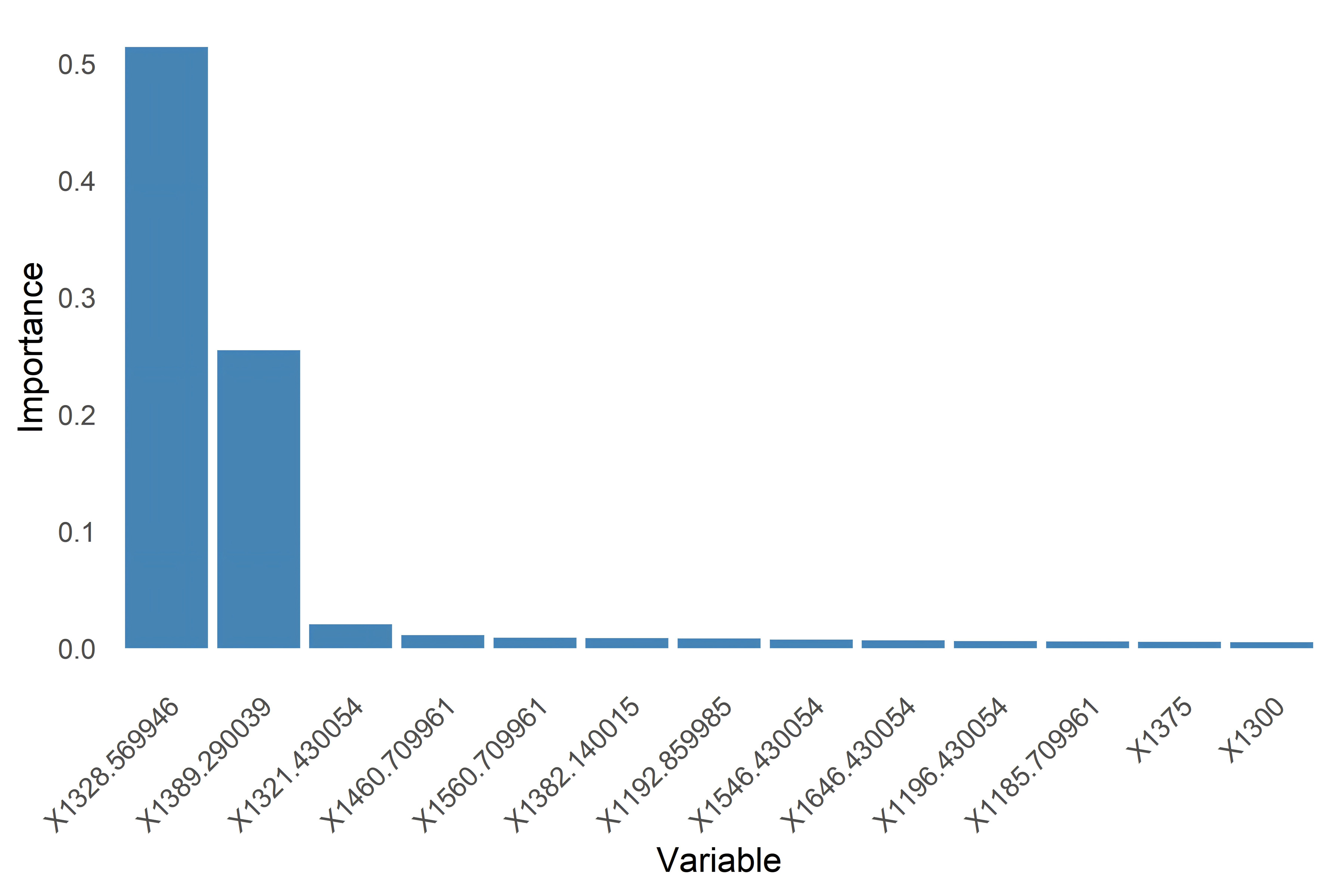}
  \caption{Posterior mean selection probabilities of variables identified as important for predicting Class 1 (pure EVOO) under the sparse Dirichlet prior. The x-axis displays the names of the selected variables, and the y-axis represents the posterior mean probability that each variable was used to split a decision node across post-burn-in MCMC iterations.}
  \label{fig:fig8}
\end{figure}

\vspace{5em}
Due to the known classification difficulty in identifying Class 1 and Class 3, the analysis here focuses on identifying variables most important for distinguishing between these two categories. Using a threshold of $1/p$ = 1/224 $\approx$ 0.0045, variables exceeding this cutoff were considered important for their respective class. Under the defined threshold of $1/p$, a total of 13 variables were identified as important for distinguishing Class 1. These are visualized in Figure~\ref{fig:fig8}.

As shown in the Figure~\ref{fig:fig8}, the posterior mean selection probabilities for the top two variables are substantially higher than the rest, which are close to or just above the threshold, indicating a clear concentration of discriminative power in a small subset of features. This suggests that, even within the selected set, a few variables dominate the model’s decision-making process for identifying Class 1. In contrast, the selection output for Class 3 includes only a single variable, X1160.709961, which exhibits a remarkably high posterior selection probability of 92.8\%, far exceeding the threshold. All other variables fall well below the cutoff, resulting in their exclusion. This asymmetry in variable selection is largely attributable to the hierarchical structure of the BART model under the multiple-binary classification setting. Specifically, the model first separates Class 1 from all other samples. Then, among the remaining observations, it distinguishes Class 2 from Class 3. As a result, the feature importance for Class 3 is conditional on the instance not being Class 2 and Class 1. This conditional framework reduces the effective sample space and may concentrate predictive power in a very limited set of features, leading to a dominant single variable being selected, while others fall below the inclusion threshold.

Given this modeling structure and the risks associated with using only one variable, the Class 3 variable selection result was excluded from further model development. The subsequent model construction was therefore based solely on the 13 selected variables for Class 1, which provided a more stable and interpretable basis for classification.

Based on the variable selection results discussed above, a new BART model was constructed using the 13 important variables identified for Class 1 classification under the sparse Dirichlet prior. These variables were used to retrain the model on the training set and subsequently evaluate its performance on the test set. 

\begin{figure}
  \centering
  \includegraphics[width=0.6\textwidth]{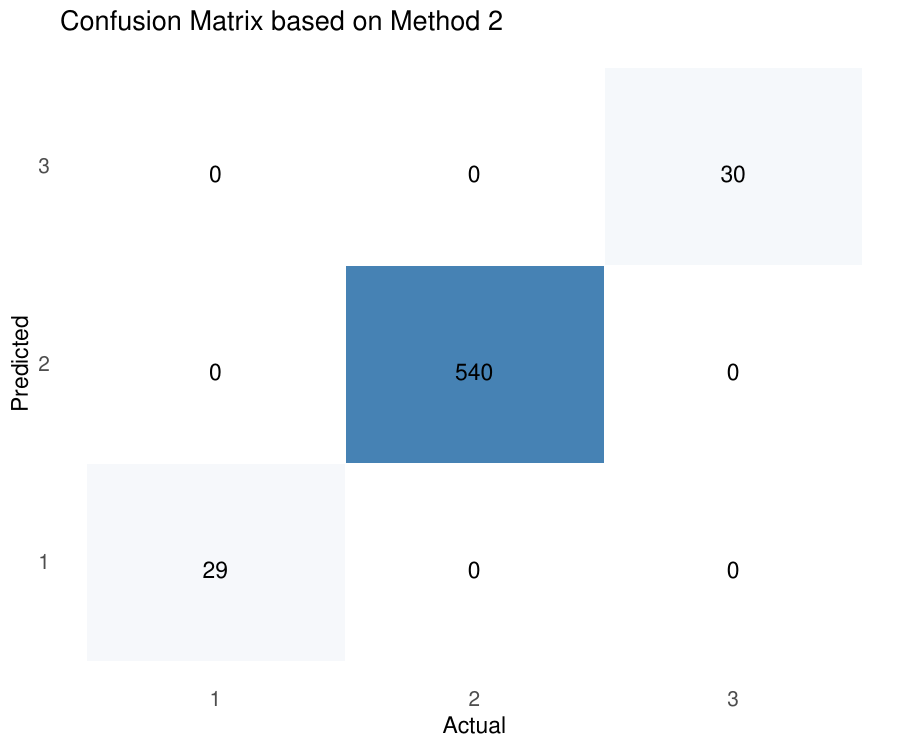}
  \caption{Confusion matrix heatmap for BART model in testing data with variables selected by method 2 (\texttt{varprob.mean}) in Class 1.}
  \label{fig:fig9}
\end{figure}

As shown in the Figure~\ref{fig:fig9}, the model achieved perfect classification across all three classes, with no misclassified samples. This result mirrors the outcome observed with the first variable selection method (based on variable usage frequency), where the model similarly reached perfect accuracy. Such consistent performance across independent variable selection strategies provides further evidence of the effectiveness and robustness of the selected feature subset, particularly in capturing the most discriminative information required for accurate classification.

Moreover, the strength of interactions between variable pairs can be shown in Figure~\ref{fig:fig11} and Figure~\ref{fig:fig12}.

\subsubsection{Variable interaction effects}
\begin{figure}
  \centering
  \includegraphics[width=0.4\textwidth]{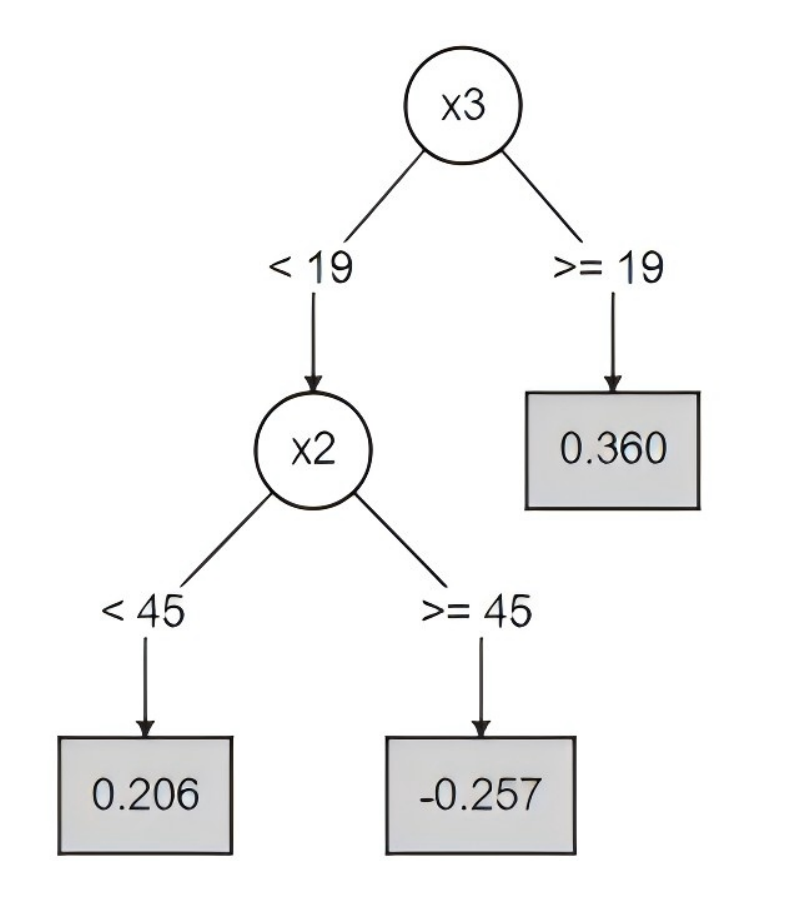}
  \caption{One of the decision trees based on the tree structure in Table 4.}
  \label{fig:fig10}
\end{figure}

To further explore the potential interaction structures among variables, a branching interaction matrix derived from the tree structures was constructed. A sample tree structure is illustrated in Table~\ref{tab:table4}, with associated tree shown in Figure~\ref{fig:fig10}.

\begin{table}[H]
 \caption{Tree structure example}
  \centering
  \begin{tabular}{c c c c}
    \toprule
    node & var & cut & leaf \\
    \midrule
    5 & NA & NA & NA \\
    1 & 2 & 19 & -0.096 \\
    2 & 1 & 45 & 0.0982 \\
    3 & 0 & 0 & 0.360 \\
    4 & 0 & 0 & 0.206 \\
    5 & 0 & 0 & -0.257 \\
    \bottomrule
  \end{tabular}
  \label{tab:table4}
  \end{table}

Table 4: Tree structure extracted from model based on method 1, where the first row represents the parent node, with all entries as NA except for node, which indicates the total number of nodes in the tree. The remaining rows represent node IDs, where cut values of 19 and 45 correspond to the decision thresholds at the first and second splits, respectively. Var values of 2 denotes to be $x_3$ and 1 denotes to be $x_2$. When both var and cut are 0, the node is a leaf node, meaning no further splitting occurs, and the fitted value leaf is output.
\vspace{5em}

The matrix is computed by traversing all trees across MCMC samples, extracting the splitting variables and cut points at each decision node. For each tree, we identify variable pairs that share the same decision path across consecutive branches and record the number of times these pairs co-occur across all trees. The matrix is then normalized by dividing the co-occurrence count of each variable pair by the total co-occurrence counts, resulting in a symmetric matrix. 

\begin{figure}
  \centering
  \includegraphics[width=0.8\textwidth]{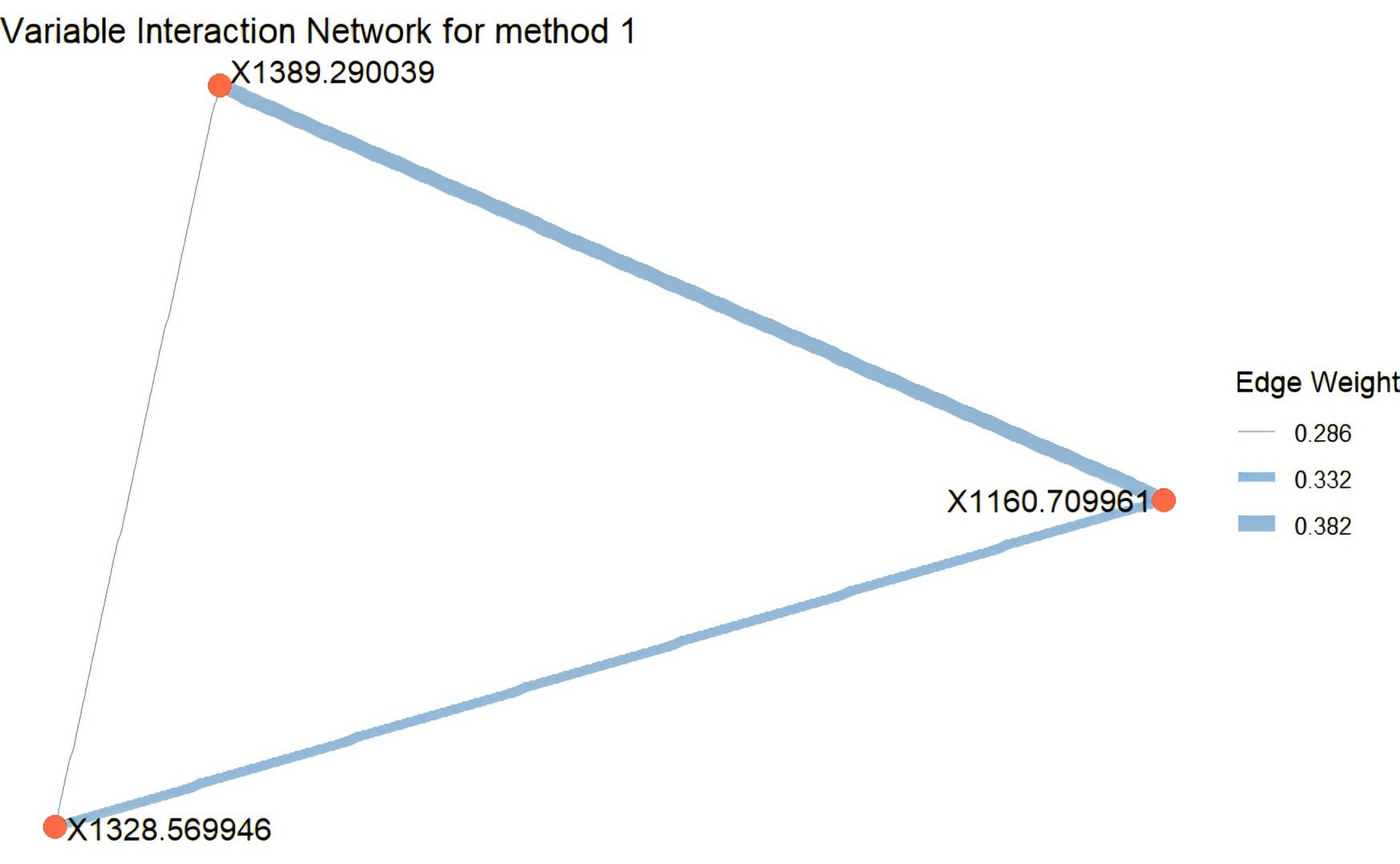}
  \caption{Interaction network to illustrate the connections between variables selected by method 1. In this network, nodes represent variables, and the edges between them are weighted to reflect the strength of their interactions. The width of the edges indicates the intensity of the interaction, with wider edges corresponding to stronger relationships between the variables.}
  \label{fig:fig11}
\end{figure}

\begin{figure}
  \centering
  \includegraphics[width=0.8\textwidth]{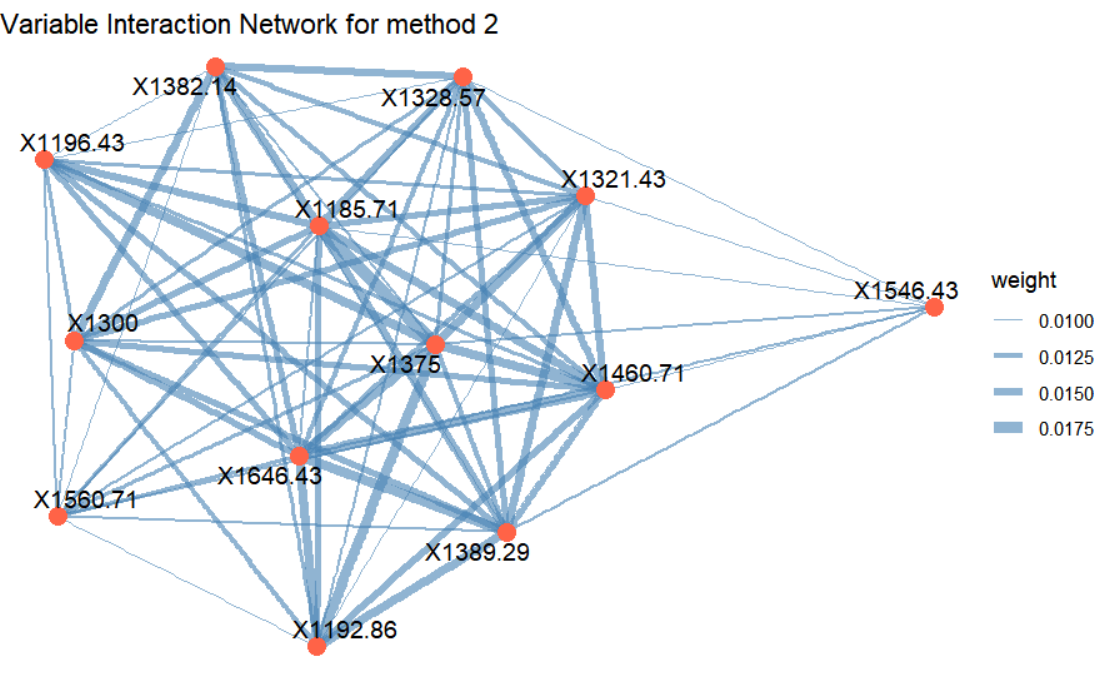}
  \caption{Interaction network to illustrate the connections between variables selected by method 2.}
  \label{fig:fig12}
\end{figure}

This matrix reflects the frequency with which each pair of variables jointly participates in the model’s decision process. Finally, the interaction patterns between the variables are visualized by using interaction networks, as shown in Figure~\ref{fig:fig11} and Figure~\ref{fig:fig12}.

Figure~\ref{fig:fig11} clearly illustrates a fully connected triangular interaction network with edge weights (variable pair co-occurrence frequencies) of 0.286, 0.332, and 0.382, which show a substantial magnitude increase compared to the results in model 2. The classification decisions in this model heavily depend on the synergistic effects among the selected feature set, with the interaction between X1389.290039 and X1160.709961 being the strongest.

By contrast, Figure~\ref{fig:fig12} depicts the second network constructed from the 13 variables selected using the second variable selection method. This network shows a more evenly distributed and larger-scale structure. The edge weights range from 0.007 to 0.019, with edges below 0.01 being filtered out in the visualization. Given that the total number of interactions exceeds 1000, edges with such low weights are considered negligible in terms of their impact on the decision-making process. Among the variables, X1375 exhibits a relatively higher degree and stronger connectivity, indicating its significant role in associating variables within the network. On the other hand, variables like X1546.430054 and X1560.71 are located at the network’s periphery with narrower edges, suggesting that these variables consistently have low interaction weights. This indicates that these variables play a more independent or marginal role in the model.

In summary, BART offers more than predictive capability. It enables structural analysis of variable relationships through its underlying decision trees. The branch matrix serves as a valuable interpretive tool to uncover latent non-linear interactions and identify feature combinations that jointly drive model performance, offering insights that are both statistically and domain-relevant.

\section{Discussion and conclusions}

Leveraging its unique ensemble tree structure and Bayesian probabilistic framework, the BART approach effectively captures nonlinear patterns and complex interactions in high-dimensional spectral data, enabling highly accurate classification of pure extra virgin olive oil, partially adulterated EVOO samples, and non-EVOO samples. Our results show that BART achieves a classification accuracy of 96.8\% under default parameter settings, comparable to traditional ensemble methods such as Random Forest. After hyperparameter tuning, the accuracy further improves to 97.2\%, demonstrating the model's outstanding predictive capability.

Through BART’s built-in variable selection mechanism, we identified some of the key wavelength that can predict the purity of olive oils. Using the variable usage frequency method, the top three selected wavelengths: 1160.71 nm, 1328.57 nm, and 1389.29 nm, were sufficient for the optimized BART model to achieve perfect classification on the test dataset. In comparison, the sparse prior-based frequency approach selected a broader subset of 13 spectral variables, most of which were concentrated between 1300 and 1400 nm, and also resulted in perfect classification performance. Notably, the three wavelengths identified through the variable usage frequency method were fully retained within the broader subset selected by the sparse prior-based approach, collectively contributing 86\% of the total classification importance in the optimized model. This finding underlines their diagnostic relevance and suggests that these wavelengths serve as the core discriminative features for effectively differentiating between pure and adulterated extra virgin olive oil samples. Specifically, the 1000–1025 nm and 1364–1421 nm regions are associated with the C–H stretching vibrations of methylene (–CH–) and methyl (–CH$_3$) groups in fatty acids \citep{xiaobo2010,malavi2024}. The spectral band around 1160.71 nm originates from the second overtone of C–H stretching vibrations related to CH$_3$ and CH$_2$ groups and their combinations \citep{xiaobo2010,tang2018,choi2020}. These wavelength regions are therefore chemically meaningful and consistent with known vibrational modes in fats and oils. Furthermore, interaction network analysis reveals that classification decisions are not made based on these variables in isolation, but rather emerge from their complex interactions within the BART model structure. This highlights the intricate and nonlinear nature of the relationships embedded in spectral data and suggests that the key chemical components corresponding to these wavelengths act synergistically to drive the classification process. Using feature selection for dimensionality reduction, the model achieves perfect classification, outperforming PCA-based dimensionality reduction methods.

These results position BART as a competitive alternative to both traditional chemometric methods and more complex deep learning approaches in food authentication. The model combines the interpretability of traditional techniques with the flexibility of modern machine learning, offering a balanced solution that addresses key challenges in spectral analysis, handling high-dimensional data, managing nonlinear relationships, and identifying the most diagnostically valuable features and their interactions. As the food industry seeks reliable and practical tools to combat economically motivated adulteration, BART emerges as a highly promising candidate solution.

Our work opens up several future research directions. One main goal would be to extend our methodology to other relevant datasets related to food fraud detections, whereby we could obtain new results and perspectives by improving upon available methods. On a different note, one practical limitation of BART is its considerable computational burden. Even with a modest set of hyperparameter combinations, a single 5-fold cross-validation can take 5–6 hours to complete when run on a standard laptop without using parallel computing. Conducting a full hyperparameter grid search may extend training time beyond 12 hours, thus highlighting the need for more scalable procedures and further research on the computational aspects of BART.
\newpage


\bibliographystyle{unsrt}



\end{document}